\documentclass[lettersize,journal]{IEEEtran}
\usepackage{amsmath,amsfonts}
\usepackage{algorithmic}
\usepackage{array}
\usepackage{subfig}
\usepackage{textcomp}
\usepackage{stfloats}
\usepackage{url}
\usepackage{verbatim}
\usepackage{graphicx}
\usepackage{pdfpages}

\def\BibTeX{{\rm B\kern-.05em{\sc i\kern-.025em b}\kern-.08em
    T\kern-.1667em\lower.7ex\hbox{E}\kern-.125emX}}
\usepackage{balance}
\usepackage{xcolor}
\usepackage{colortbl}
\definecolor{amber}{rgb}{1.0, 0.75, 0.0}
\definecolor{ao(english)}{rgb}{0.0, 0.5, 0.0}
\usepackage{upgreek} 
\usepackage{array}
\usepackage{cuted}
\newcolumntype{P}[1]{>{\centering\arraybackslash}p{#1}}
\usepackage{multirow}

\usepackage{verbatimbox}
\usepackage{atbegshi}

\begin{document}
\AtBeginShipoutNext{%
  \AtBeginShipoutUpperLeft{%
    \put(\dimexpr\paperwidth-21cm\relax,-0.5cm){%
      \makebox[0pt][l]{{\tiny This article has been accepted for publication in IEEE Transactions on Circuits and Systems--II: Express Briefs. This is the author's version which has not been fully edited and
content may change prior to final publication. Citation information: DOI 10.1109/TCSII.2023.3302235  }}%
    }%
  }%
}

\title{Experimental Demonstration of Non-Stateful In-Memory Logic with 1T1R OxRAM Valence Change Mechanism Memristors \\}

\author{Henriette Padberg, Amir Regev, Giuseppe Piccolboni, Alessandro Bricalli, Gabriel Molas, \textit{Senior Member, IEEE}, Jean Francois Nodin, and Shahar Kvatinsky, \textit{Senior Member, IEEE}%IEEE Publication Technology,~\IEEEmembership{Staff,~IEEE,}% <-this % stops a space
\thanks{Manuscript received 1 May 2023; revised 9 July 2023; accepted 2 August 2023. This work was supported by the European Research Council through the European Union’s Horizon 2020 Research and Innovation Programme under Grant 757259 and through the European Union’s Horizon Europe Research and Innovation Programme under Grant 101069336. This brief
was recommended by Associate Editor X. Miao. (Corresponding author:
Henriette Padberg.)}%<-this % stops a space
\thanks{H. Padberg and S. Kvatinsky are with the Department of Electrical Engineering,
Technion Israel Institute of Technology, Haifa 32000, Israel \\(e-mail: henriette.padberg@rwth-aachen.de; shahar@ee.technion.ac.il).}%
\thanks{Amir Regev is with Weebit Nano, 4527713 Hod Hasharon, Israel (e-mail: amir@weebit-nano.com).}

\thanks{Giuseppe Piccolboni, Alessandro Bricalli, and Gabriel Molas are with Weebit Nano, 38000 Grenoble, France.}
\thanks{J. F. Nodin is with CEA/Leti, Grenoble, France.}
\thanks{Color versions of one or more figures in this article are available at
https://doi.org/10.1109/TCSII.2023.3302235 .} 
\thanks{Digital Object Identifier 10.1109/TCSII.2023.3302235} 
}%

\markboth{IEEE TRANSACTIONS ON CIRCUITS AND SYSTEMS—II: EXPRESS BRIEFS}%
{Shell \MakeLowercase{\textit{et al.}}: A Sample Article Using IEEEtran.cls for IEEE Journals}

\IEEEpubid{\begin{minipage}{\textwidth}\ \\[32pt] \centering
  1549-7747 \copyright 2023 IEEE. Personal use is permitted, but republication/redistribution requires IEEE permission.\\
  See http://www.ieee.org/publications standards/publications/rights/index.html for more information.
\end{minipage}}
%\IEEEpubid{0000--0000/00\$00.00~\copyright~2023 IEEE}

\maketitle

\begin{abstract}
Processing-in-memory (PIM) is attractive to overcome the limitations of modern computing systems. Numerous PIM systems exist, varying by the technologies and logic techniques used. Successful operation of specific logic functions is crucial for effective processing-in-memory. Memristive non-stateful logic techniques are compatible with CMOS logic and can be integrated into a 1T1R memory array, similar to commercial RRAM products. This paper analyzes and demonstrates two non-stateful logic techniques: 1T1R logic and scouting logic. As a first step, the used 1T1R SiO\textsubscript{x} valence change mechanism memristors are characterized in reference to their feasibility to perform logic functions. Various logical functions of the two logic techniques are experimentally demonstrated, showing correct functionality in all cases. Following the results, the challenges and limitations of the RRAM characteristics and 1T1R configuration for the application in logical functions are discussed. 

\end{abstract}

\begin{IEEEkeywords}
1T1R Logic, Non-Stateful Logic, Scouting Logic, Experimental Demonstration, Reliability Issues
\end{IEEEkeywords}

\section{Introduction}

\IEEEPARstart{I}{n} times of reaching the end of Moore's law and dealing with the issue of the memory wall, the search for efficient methods of non-von Neumann computing employs many scientists and engineers~\cite{Theis.2017,Williams.2017}. Recently, the research fields of neuromorphic computing and computing in-memory have caught a lot of attention~\cite{Vishwa.2020, Ou.2022, Ielmini.2018}, {\cite{Sun.2018}}. Computing in-memory implies that logic operations are performed directly in the memory without costly data transfer between the memory and a separate processing unit \cite{Kvatinsky.2019}. Such computing promises energy-efficient computing with the potential to overcome the von Neumann bottleneck. Computing in-memory can be realized with nonvolatile devices with the resistive random access memory (RRAM) as an outstanding candidate due to its various advantages in power consumption, speed, durability, and compatibility for 3D integration \cite{Li.2017, Waser.2012}.
One promising approach for RRAM-based computing is stateful logic such as MAGIC \cite{Kvatinsky.2014} and IMPLY \cite{Borghetti.2010}. In stateful logic, the logical states of inputs and outputs are represented as the resistance states of the memristor devices, with logical '0' as a High Resistance State (HRS) and logical '1' as a Low Resistance State (LRS). However, stateful logic techniques have yet to be demonstrated for large-scale crossbar array implementation. Furthermore, stateful logic is incompatible with CMOS logic and is limited by the device's endurance \cite{Resch.2023}. Another approach is non-stateful logic, in which different electrical variables represent their inputs and outputs. For example, the inputs are voltages, and the output is the resistance state of the memristor. Non-stateful logic combines the advantages of computing in-memory and CMOS compatibility.

Commercial RRAM products are built in a 1T1R configuration, where every memory cell has a transistor and a memristive device. Wang et al.~\cite{Wang.2017} proposed a functionally-complete Boolean logic based on 1T1R arrays by defining the parameters for the voltages at the gate of the transistor, $G$, the top electrode, $TE$, and the bottom electrode, $BE$, of the memristor as the inputs of the logic gate. Another input is the memristor's initial resistive state, $I$. The output of the logic gate is read out as the resistive state of the memristor after the logic operation. All four parameters ($G$, $TE$, $BE$, and $I$) are defined according to the values of the logic gate's inputs, $p$ and $q$, with four possible combinations. In this brief, the described logic will be referred to as \textit{1T1R logic}.    

Another non-stateful logic type suitable for 1T1R arrays is \textit{scouting logic}~\cite{Xie.2017}. Here, the inputs are represented by the resistive states of two memristors, and the output is the measured current during a simultaneous read-out of the inputs. Depending on the selection of a reference current, the logical functions AND, OR, and XOR are performed. 

This brief experimentally demonstrates 1T1R as well as scouting logic techniques using SiO\textsubscript{x}-based Valence Change Mechanism (VCM) memristors to explore their possibilities for various applications. We show successful operations of both logic types and highlight the critical failure risks, such as the overlap of HRS and LRS state and state-instabilities. Furthermore, the limitations of the 1T1R configuration for other logic techniques are discussed. The results confirm the reliability issues of RRAM devices, especially variability, which is one of the greatest challenges of VCM cells.

\section{Experimental Measurements}

\subsection{Sample}
\begin{figure*}[!t]
\centering
\subfloat[]{\scalebox{0.7}{\includegraphics[width=0.27\textwidth]{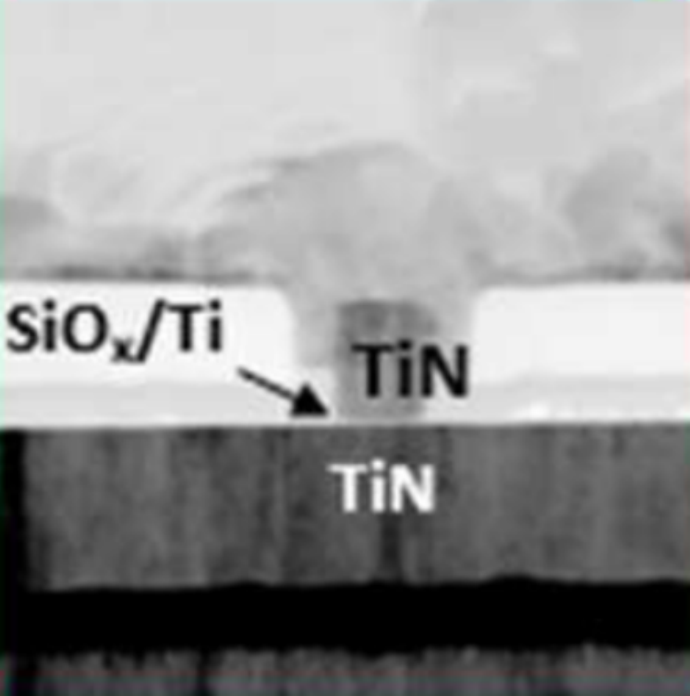}}}
\hfil
\subfloat[]{\scalebox{1.0}{\includegraphics[width=0.27\textwidth]{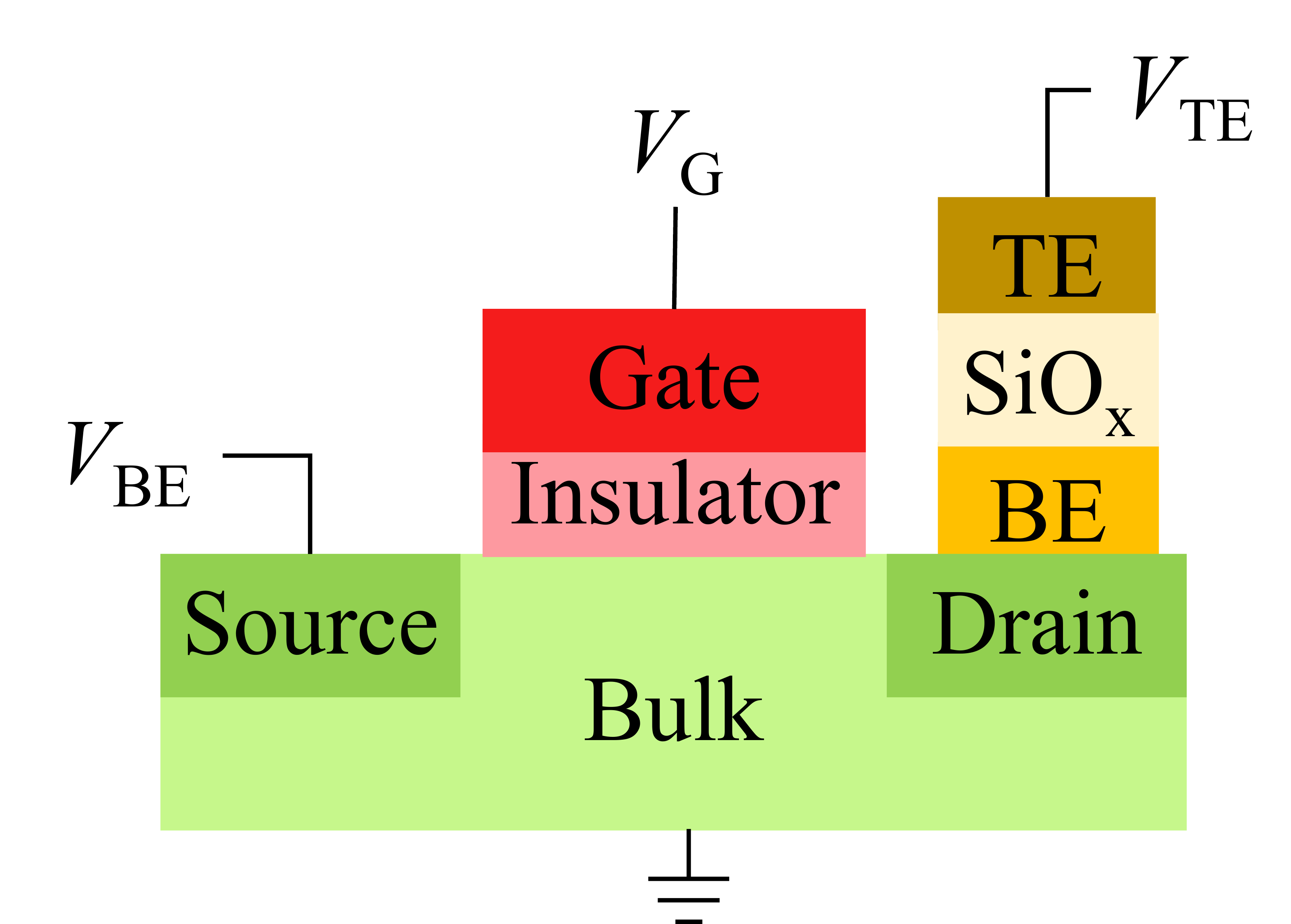}}}
\hfil
\subfloat[]{\scalebox{0.7}{\includegraphics[width=0.27\textwidth]{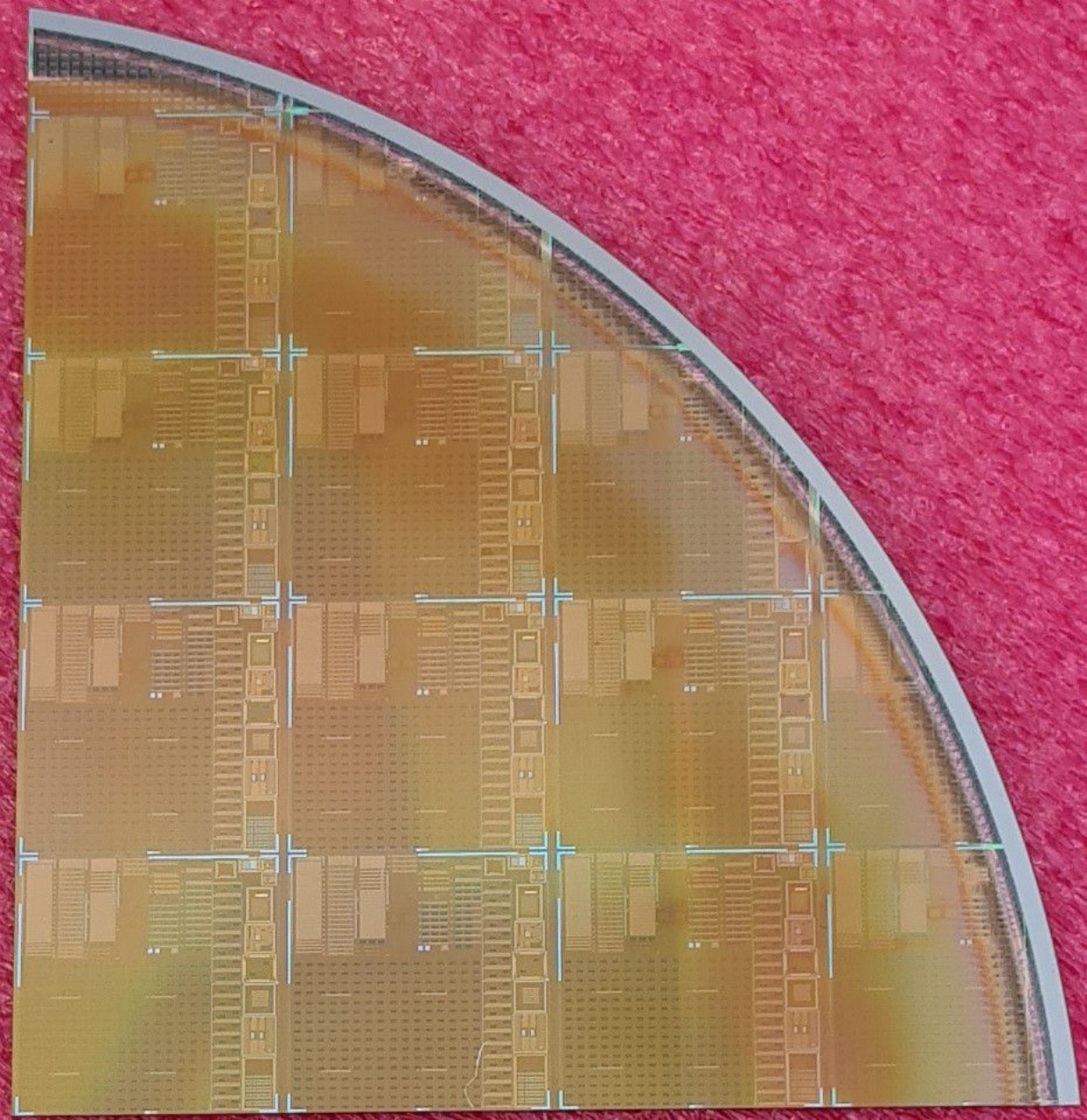}}}
\\
\subfloat[]{\scalebox{0.78}{\includegraphics[width=0.3\textwidth]{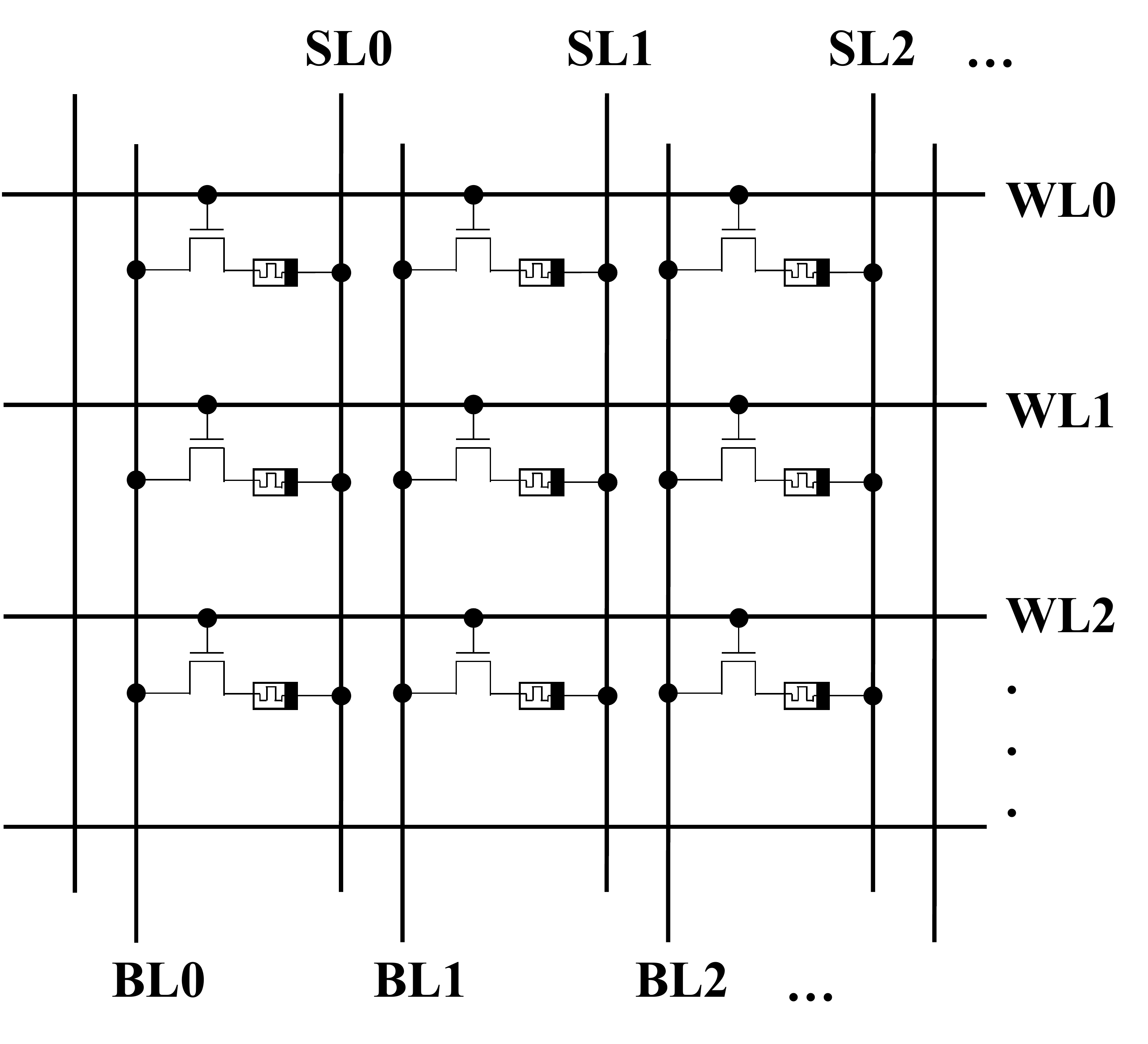}}}
\hfil
\subfloat[]{\scalebox{1}{\includegraphics[width=0.3\textwidth]{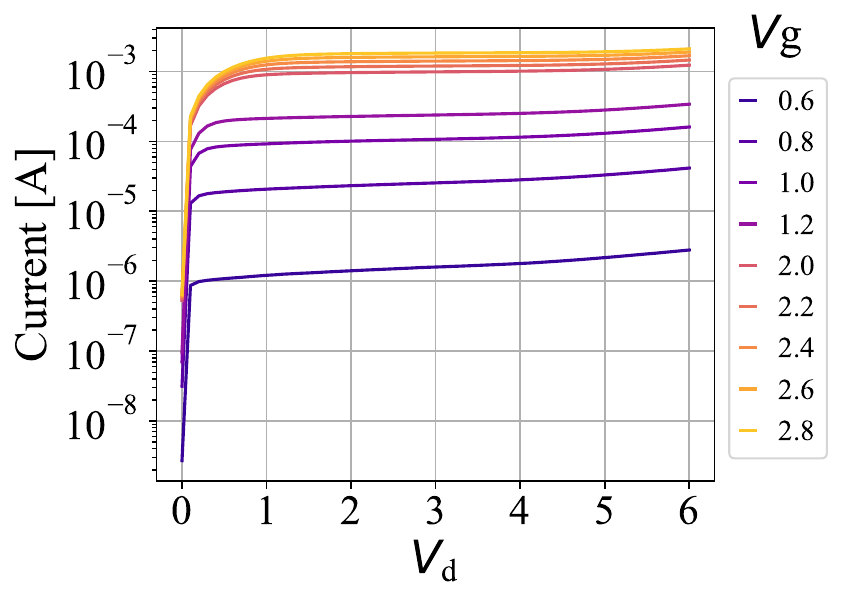}}}
\hfil
\subfloat[]{
    \label{table:input_possibilities}
	\renewcommand{\arraystretch}{1.2}
 \resizebox{0.24\textwidth}{!}{%
        \begin{tabular}[b]{c | c |c |c |c }
        
        \hline \hline
        
        \rowcolor{gray!20} & \textbf{Parameter}& \textbf{\textcolor{ao(english)}{Nominal}} & \textbf{\textcolor{amber}{Min}}  &  \textbf{\textcolor{red}{Max}} \\
         \hline
        \multirow{3}{*}{\rotatebox{90}{Forming}} & $V_{TE}$\,[V] & 0:0.1:4.8 & NA & NA\\
        &$t_\mathrm{pulse}$\,[s] &10\textsuperscript{-5} & 3$\cdot$10\textsuperscript{-5} & NA \\
        & $V_\mathrm{Gate}$\,[V] & 1.1 & NA & NA \\
        \hline
        \multirow{3}{*}{\rotatebox{90}{SET}} & $V_{TE}$\,[V] & 2 & 1.8 & 2.3\\
        &$t_\mathrm{pulse}$\,[s] & 10\textsuperscript{-6} & 3$\cdot$10\textsuperscript{-7} & NA \\
        & $V_\mathrm{Gate}$\,[V] & 1.1 & NA & NA \\
        \hline
        \multirow{3}{*}{\rotatebox{90}{RESET}} & $V_{TE}$\,[V] & 1.2 & 1.0 &1.5\\
        &$t_\mathrm{pulse}$\,[s] &10\textsuperscript{-5}/10\textsuperscript{-4} & 3$\cdot$10\textsuperscript{-6} & NA \\
        & $V_\mathrm{Gate}$\,[V] & 4.5 & NA & NA \\
        \hline \hline
        \end{tabular}}
}%

\caption{VCM sample structure and configuration. (a) SEM image of the MIM structure of the Weebit-Nano VCM cells~\cite{Goes.2021}. (b) Scheme of the 1T1R stack. (c) Image of the full sample. (d) Schematic of the  1T1R memory array. (e) The transistor characteristics by current over drain voltages $V_d$ for different gate voltages $V_g$. (f) Ideal switching parameters of the VCM devices \cite{Piccolboni.2019}.}
\label{fig:sample}
\end{figure*}

\begin{figure}[t!]
\centering
\subfloat[]{\scalebox{0.7}{\includegraphics[width=0.5\textwidth]{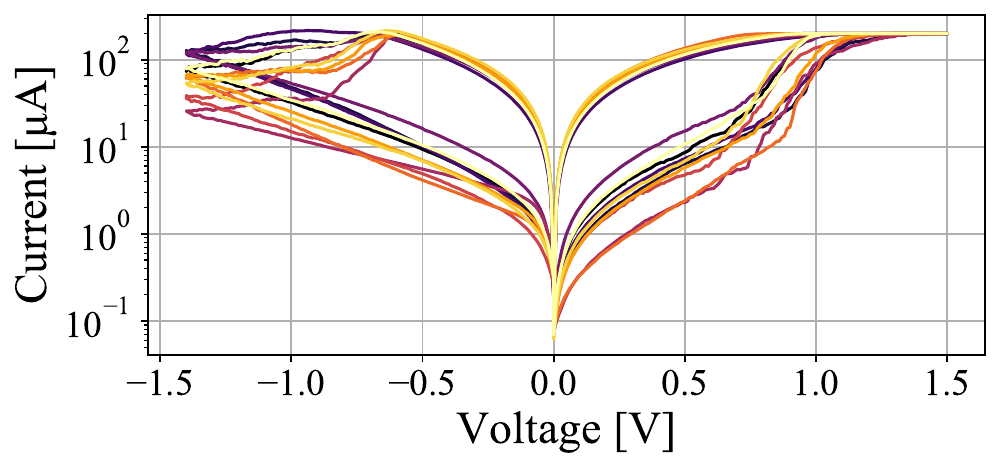}}%
}
\hspace{3cm}
\\
\subfloat[]{\scalebox{0.8}{\includegraphics[width=0.5\textwidth]{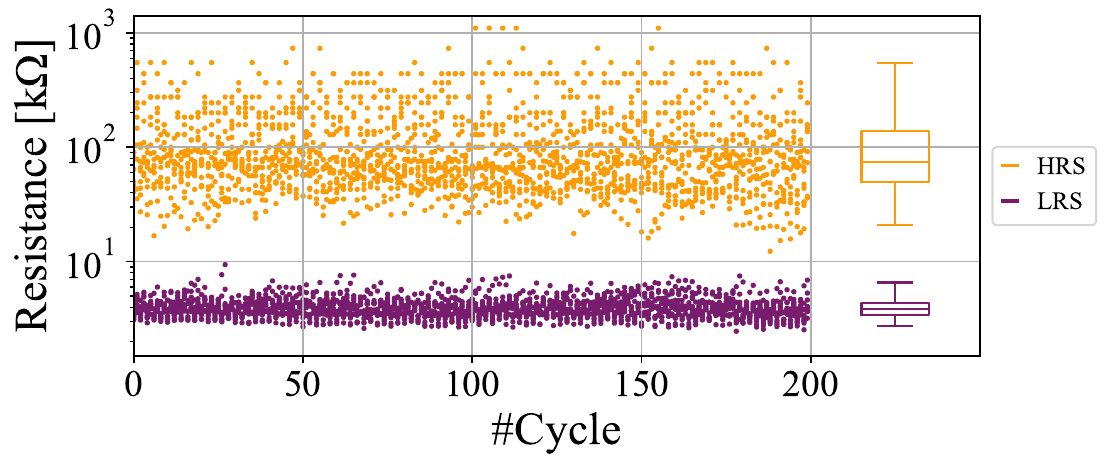}}%
}\hfill

\caption{(a) I-V curves of ten different 1T1R cells. (b) Resistances of HRS and LRS extracted from 100 cycles of 10 cells, 1\%-99\%-whisker box plot.}
\label{fig:cell characteristics}
\end{figure}

The measurements were performed using the metal-insulator-metal (MIM) structure of VCM cells provided by Weebit-Nano, fabricated at CEA/Leti R\&D center in Grenoble France \cite{Regev.2020}. The devices have the stack composition of TiN as the bottom electrode (BE), SiO\textsubscript{x} as the switching material, and Ti as the top electrode (TE). Further device information is shown in Fig.~\ref{fig:sample}. 
The cell performs a SET process when a positive voltage is applied at the TE with the BE grounded and performs a RESET with a positive voltage at the BE, and the TE is connected to the ground. The switching characteristics of ten different cells are shown in Fig.~\ref{fig:cell characteristics}, demonstrating cell-to-cell variability and the cycle-to-cycle variability of HRS and LRS. Nevertheless, the resistances of HRS and LRS do not overlap at any time, with a mean HRS/LRS ratio of 19.4, ensuring a clear distinction between HRS and LRS.

\subsection{Measurement Setup}
All tests were performed using a cascade summit12000 probe station, controlled by a Keysight B1500A parameter analyzer. 

\subsubsection{1T1R Logic}
To realize the combinations of the input parameters for the 16 possible Boolean functions \cite{Wang.2017}, $G$ is connected to the WL, $TE$ is linked to the SL, $BE$ is connected to the BL, and an additional needle grounds the bulk of the transistor. The WL, SL, and BL are defined according to the array structure shown in Fig.~\ref{fig:sample} (c).

Since there are four input parameters, there are 16 possible input combinations. These configurations are listed in Table~\ref{tab:input_pos}. The column '$TE$-$BE$' shows if there is a voltage drop over the cell, it is +1 for a positive voltage, -1 for a negative voltage, and 0 for no voltage difference. According to the definition of the cell, a SET is performed for a positive voltage and a RESET for a negative voltage, so the resulting process for $TE$-$BE$=+1 is a SET and for $TE$-$BE$=-1 a RESET. If $TE$-$BE$=0, no process can happen due to the lack of a voltage difference. However, not in all cases of Table~\ref{tab:input_pos} where $TE$-$BE$=+1/-1 the execution of the given process is possible since the values of $G$ and $I$ have to be considered as well. The transistor is closed if $G$=0 (cases 11-14); therefore, switching is impossible in these cases.  Furthermore, if the initial state of the memristor $I$ is 1 and the given process is a SET (case 3), $I$ is already in the LRS so that no switching will occur. This also counts for $I$=0 in combination with a RESET process (case 6). Therefore, all four inputs enable the given switching process only in cases 4 and 5.

\begin{table}[t!]
	\caption[]{Possible combinations of $G$, $TE$, $BE$, and $I$.}
    \label{tab:input_pos}
	\centering
	\renewcommand{\arraystretch}{1}
\begin{tabular}{c | c c c c | c |P{1.6cm} }
\hline \hline

\rowcolor{gray!20}\multirow{2}{*}{\textbf{case}} & \multirow{2}{*}{\textbf{G}}&  \multirow{2}{*}{\textbf{TE}} &  \multirow{2}{*}{\textbf{BE}}  &   \multirow{2}{*}{\textbf{I}}  &  \multirow{2}{*}{\textbf{TE-BE}} & \textbf{Process} (\textbf{Possible?}) \\
 \hline
1) & 1 & 1 & 1 &  1 & 0 & / 
\\

2) & 1 & 1 & 1 & 0 & 0 & /                   \\

3) & 1 & 1 & 0  & 1 & +1 & SET  (no)     \\

\textbf{4)} & \textbf{1} & \textbf{1} & \textbf{0} & \textbf{0}  & \textbf{+1} & \textbf{SET  (yes)}     \\

\textbf{5)} & \textbf{1}  & \textbf{0} & \textbf{1} &  \textbf{1}  & \textbf{-1} & \textbf{RESET   (yes)  }       \\

6) & 1  & 0 & 1 &  0 & -1 & RESET (no)     \\

7) & 1 & 0  & 0 & 1 & 0 & /                   \\

8) & 1 & 0  & 0 &  0 & 0  & /            \\

9) & 0 & 1 & 1 & 1 & 0 & /  \\

10) & 0 & 1 & 1 & 0 & 0 & /  \\

11) & 0 & 1 & 0 & 1 & +1 & SET  (no) \\

12) & 0 & 1 & 0 & 0 & +1  & SET  (no) \\

13) & 0 & 0 & 1 & 1 & -1 & RESET  (no) \\

14) & 0 & 0 & 1 & 0 & -1 & RESET  (no)\\

15) & 0 & 0 & 0 & 1  & 0 & /  \\

16) & 0 & 0 & 0 & 0 & 0 & /  \\
\hline \hline
\end{tabular}
\end{table}

Given a truth table with the inputs $p$ and $q$, each logic function can be realized by setting the parameters $G$, $TE$, $BE$, and $I$ to constants or the inputs $p$ and $q$~\cite{Wang.2017}. Each combination of $G$, $TE$, $BE$, and $I$ equals one of the cases in Table~\ref{tab:input_pos}, and for each logic function, four cases are possible out of the 16 configurations.
For example, if the inputs are $p$=$q$=1 in the configuration of the NOT $p$ function ($G=0$, $TE=0$, $BE=q$, and $I=\overline{p}$), the resulting input combination equals case 14 of Table~\ref{tab:input_pos}.

Similarly, we tested four logic functions (Table~\ref{table:logic_experiment}): OR, AND, NIMP, and XOR. 
These four functions were chosen because of their importance for different applications. Furthermore, these functions include cases 3-8, 12, 13 and 16, covering all potentially critical cases ($TE-BE$ $\neq$0) and some uncritical cases with $TE-BE$ = 0. The cases out of Table~\ref{tab:input_pos} not included by these functions can always be considered non-critical for a correct logical output since there is no possible logical failure without a voltage difference between TE and BE. The cases 11 and 14 do have a voltage difference, however, the transistor is closed and additionally, the voltage polarity is leading to the opposite switching process, so that also these two cases can be considered as non-critical.

\begin{table}[!t]
	\caption[]{Input Parameters for four logic operations.}
	\centering
	\label{table:logic_experiment}
	\renewcommand{\arraystretch}{1}
\begin{tabular}{P{1cm}|P{0.125cm} P{0.14cm} P{0.14cm} P{0.135cm}|P{0.35cm}||P{1cm}|P{0.125cm} P{0.14cm} P{0.14cm} P{0.135cm}|P{0.3cm}}
\hline \hline

\rowcolor{gray!20}& \textbf{G} &\textbf{TE} &\textbf{BE} & \textbf{I} & \textbf{Out} & 
& \textbf{G} & \textbf{TE} & \textbf{BE} & \textbf{I} & \textbf{Out}  \\

\rowcolor{gray!20} \textbf{OR} & 1 & \textcolor{brown}{$q$} & 0 & \textcolor{brown}{$p$} &
&\textbf{AND} & \textcolor{brown}{$p$} & \textcolor{brown}{$q$} & 0 & 0 &\\
\hline

a) $\rightarrow$ 8 & 1 & \textcolor{brown}{0} & 0 & \textcolor{brown}{0} & 0 &
a) $\rightarrow$ 16 & \textcolor{brown}{0} & \textcolor{brown}{0} & 0 & 0 & 0 
\\ 

b) $\rightarrow$ 4 & 1 & \textcolor{brown}{1} & 0 & \textcolor{brown}{0} & 1 &
b) $\rightarrow$ 12 & \textcolor{brown}{0} & \textcolor{brown}{1} & 0 & 0 & 0 
\\

c) $\rightarrow$ 7 & 1 & \textcolor{brown}{0} & 0 & \textcolor{brown}{1} & 1 &
c) $\rightarrow$ 8 & \textcolor{brown}{1} & \textcolor{brown}{0} & 0 & 0 & 0    
\\

d) $\rightarrow$ 3  & 1 & \textcolor{brown}{1} & 0 &\textcolor{brown}{1} & 1  &
d) $\rightarrow$ 4 & \textcolor{brown}{1} & \textcolor{brown}{1} & 0 & 0 & 1 
\\

\hline\hline

\rowcolor{gray!20} & \textbf{G} & \textbf{TE} & \textbf{BE} & \textbf{I} & \textbf{Out} &  
& \textbf{G} & \textbf{TE} & \textbf{BE} & \textbf{I} & \textbf{Out}  \\
\rowcolor{gray!20} \textbf{NIMP} & 1 & 0 & \textcolor{brown}{$p$} & \textcolor{brown}{$q$} &
& \textbf{XOR} & \textcolor{brown}{$q$} & \textcolor{brown}{$\overline{p}$} & \textcolor{brown}{$p$} & \textcolor{brown}{$p$} &\\
\hline

a) $\rightarrow$ 8 & 1 & 0 & \textcolor{brown}{0} & \textcolor{brown}{0} & 0 &
a) $\rightarrow$ 12 & \textcolor{brown}{0} & \textcolor{brown}{1} & \textcolor{brown}{0} & \textcolor{brown}{0} & 0 
\\ 

b) $\rightarrow$ 7 & 1 & 0 & \textcolor{brown}{0}& \textcolor{brown}{1} & 1 &
b) $\rightarrow$ 4 & \textcolor{brown}{1} & \textcolor{brown}{1} & \textcolor{brown}{0} & \textcolor{brown}{0} & 1 
\\

c) $\rightarrow$ 6 & 1 & 0 & \textcolor{brown}{1} & \textcolor{brown}{0} & 0 &
c) $\rightarrow$ 13 & \textcolor{brown}{0} & \textcolor{brown}{0} & \textcolor{brown}{1} & \textcolor{brown}{1} & 1 
\\

d) $\rightarrow$ 5 & 1 & 0 & \textcolor{brown}{1} & \textcolor{brown}{1} & 0 &
d) $\rightarrow$ 5 & \textcolor{brown}{1} & \textcolor{brown}{0} & \textcolor{brown}{1} & \textcolor{brown}{1} & 0 
\\
\hline \hline
\end{tabular}
\label{tab:1T1R logic cases}
\end{table}

\subsubsection{Scouting Logic}
\begin{figure}[t!]
\centering
\subfloat[]{\scalebox{1}{\includegraphics[width=0.17\textwidth]{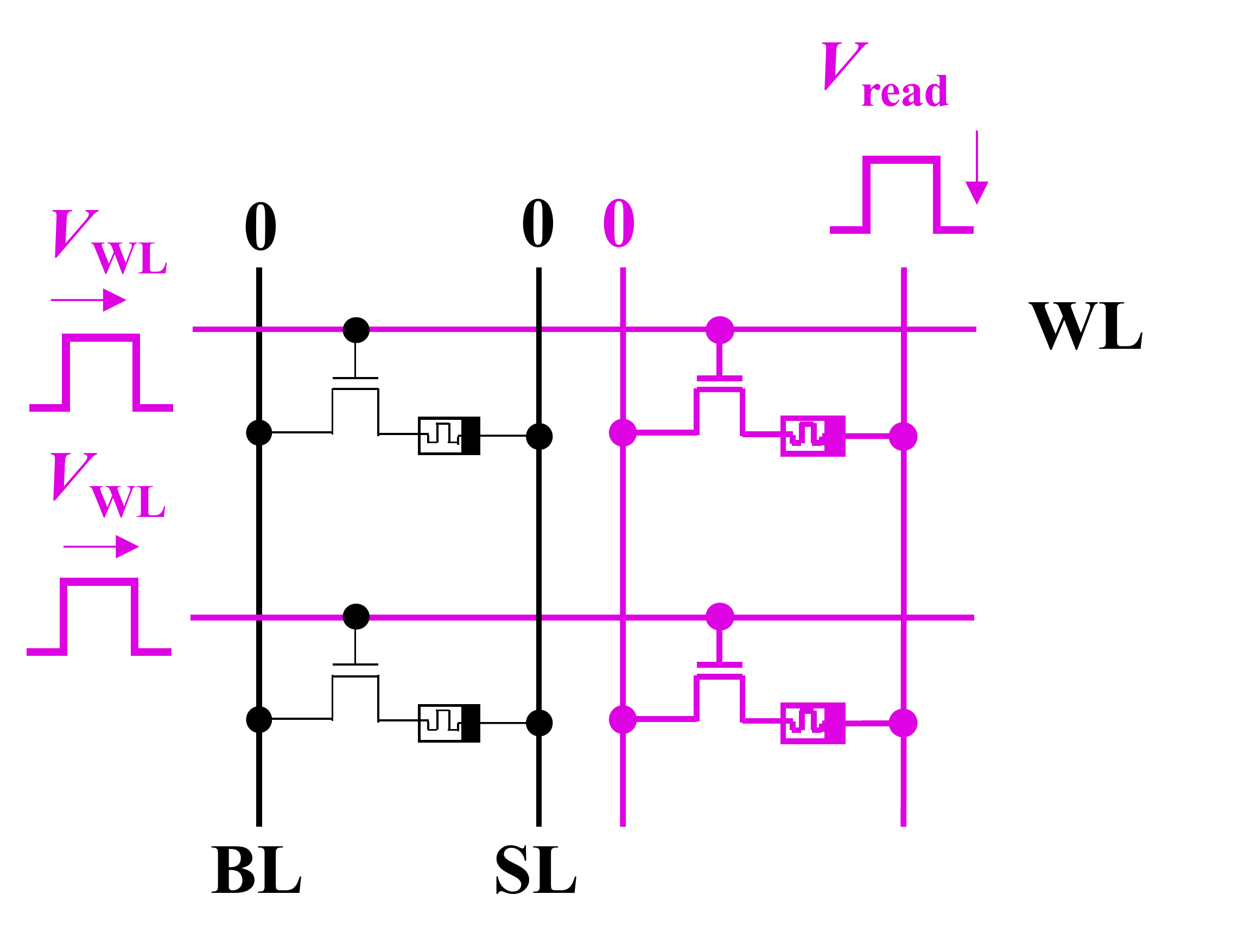}}%
}%
\subfloat[]{\scalebox{1}{\includegraphics[width=0.33\textwidth]{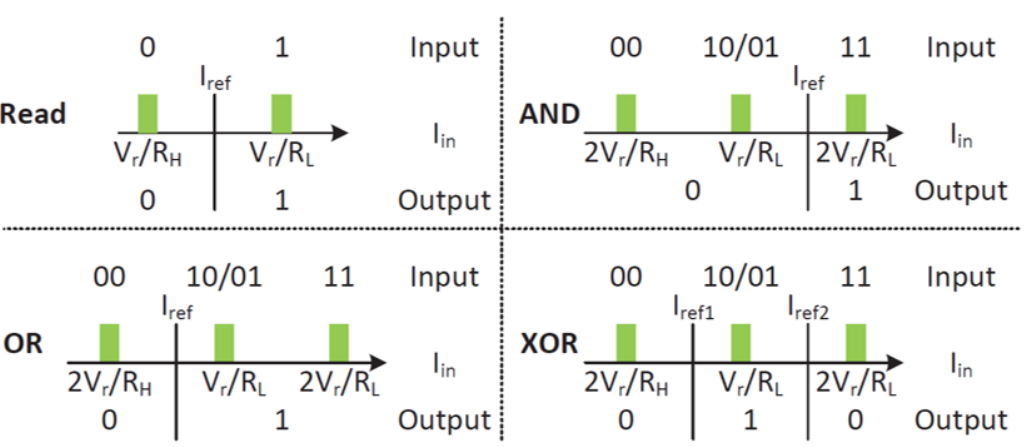}}%
}

\caption{(a) Array connection to perform scouting logic and (b) selection of reference currents for scouting logic operations~\cite{Xie.2017}.}
\label{fig:Scouting principle}
\end{figure}

In scouting logic, the inputs are represented by the resistive states of two or more memristors, and the output is the current through the memristors while applying a read-voltage~\cite{Xie.2017}. The assignment of the current to logical '1' or '0' is done by defining a reference current. Current higher (lower) than the reference current is defined as '1' ('0'). Fig.~\ref{fig:Scouting principle}(b) shows the definition of reference currents for Read, AND, OR, and XOR functions. The input combination 10 (cell 1: LRS, cell 2: HRS) is at the same position as 01 (cell 1: HRS, cell 2: LRS) since, in both cases, one memristor is in the HRS, the other in the LRS, and the current should be therefore similar. Two reference currents are needed for the XOR function because the input combinations 00 and 11 have the same output '0'.

For the experimental realization of scouting logic, a read voltage is applied at two parallel connected cells, and the resulting current is measured;$V_\mathrm{read}$ is applied at one SL and $V_\mathrm{WL}$ at two WLs to open two transistors, as shown in Fig.~\ref{fig:Scouting principle}(a).

Each measurement for both 1T1R and scouting logic is performed with the measurement parameters listed in Table~\ref{tab:meas parameter}. The measurement for each logic gate is repeated 100 times to consider cycle-to-cycle variability and was executed in a cascading structure. This was realized by reading out the state of the corresponding memristor after the logic operation and an additional initialization if the cell is not in the logical state that is required for the next cascade.

\begin{table}[!t]
\caption[]{Measurement parameters for both logic types.}
\centering
\label{tab:meas parameter}
\renewcommand{\arraystretch}{1.0}

    \begin{tabular}{c | c | c | c}
    \hline \hline
      \rowcolor{gray!20}\textbf{variable}& \textbf{SET} & \textbf{RESET}  &  \textbf{read} \\

    \hline
      $V_{\mathrm{TE}}$ [V] & 1.3 & 0 & 0.1 \\
            $V_{\mathrm{BE}}$ [V]& 0 & 1.6 & 0 \\
            $V_{\mathrm{G}}$ [V] & 1.3 & 3 & 3 \\
            $t_{\mathrm{pulse}}$ [$\upmu$s] & 1 & 1 & 1\\

\hline \hline
\end{tabular}
\end{table}

\section{Results and Discussion}
\subsection{Complete Boolean Logic with 1T1R}

\begin{figure}[t!]
\centering
\subfloat[]{\scalebox{1}{\includegraphics[width=0.24\textwidth]{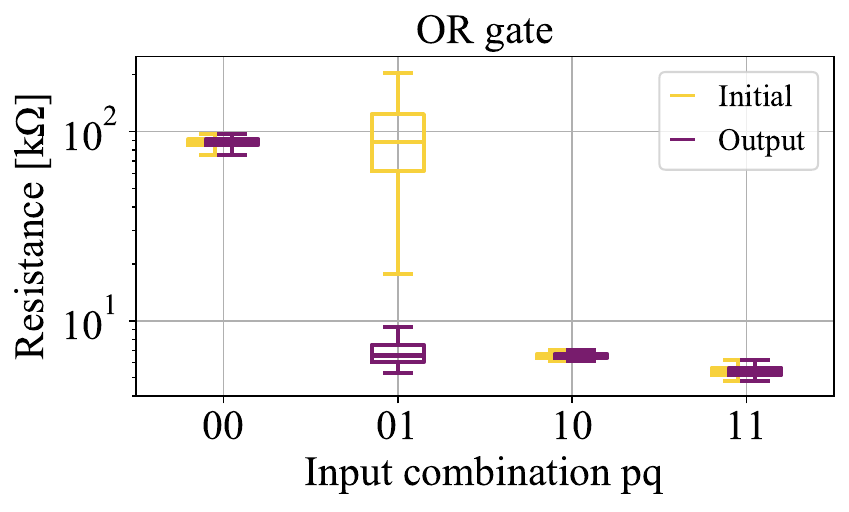}}%
}
\hfill
\subfloat[]{\scalebox{1}{\includegraphics[width=0.24\textwidth]{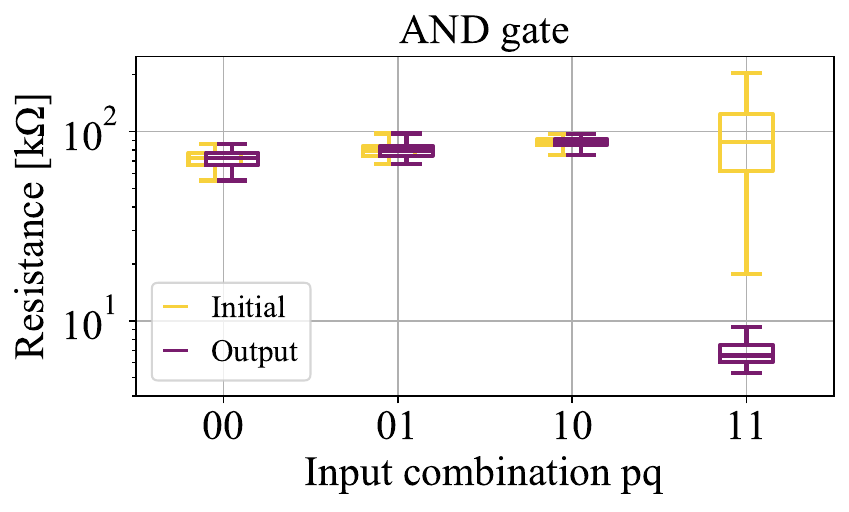}}%
}
\\
\subfloat[]{\scalebox{1}{\includegraphics[width=0.24\textwidth]{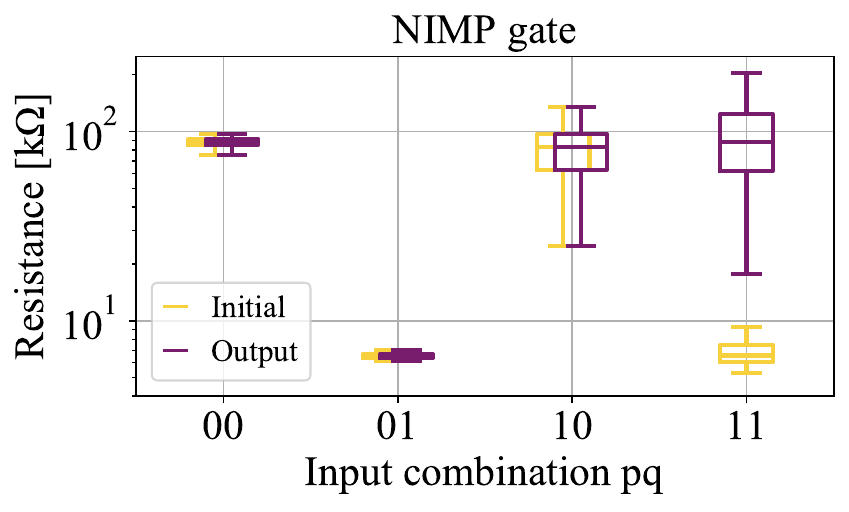}}%
}
\hfill
\subfloat[]{\scalebox{1}{\includegraphics[width=0.24\textwidth]{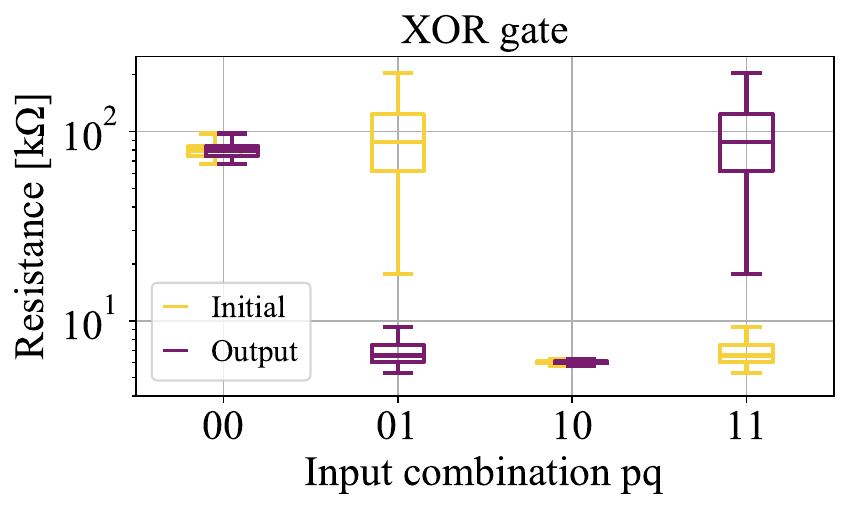}}%
}
\\
\subfloat[]{\scalebox{1}{\includegraphics[width=0.45\textwidth]{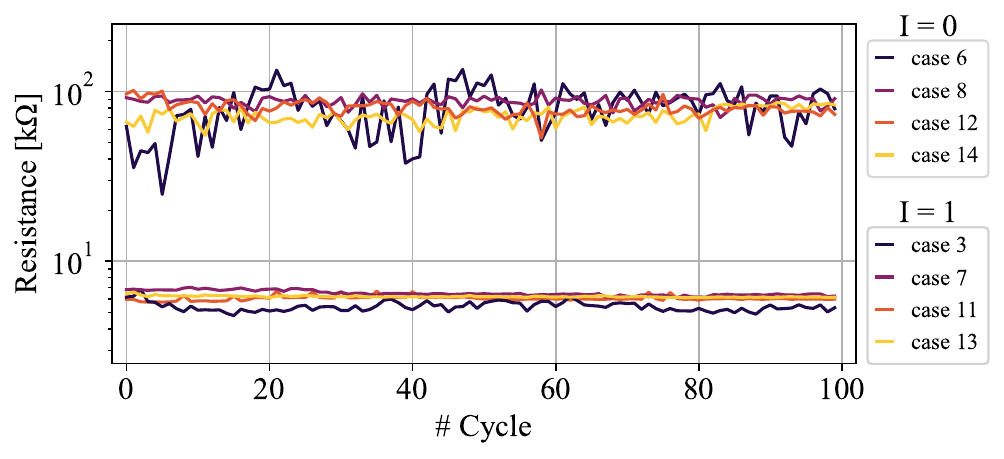}}%
}
\caption{Experimental results for (a)~OR, (b) AND, (c) NIMP, and (d) XOR logic gates. (e) Resistance for all non-switching cases.}
\label{fig:results gates}
\end{figure}

Fig.~\ref{fig:results gates} shows the results of four logic gates for all four input combinations. The required switching in cases 4 and 5 are observed in all of the logic functions, demonstrating successful switching in each cycle. However, the HRS distribution ranges over an order of magnitude, implying a relatively high cycle-to-cycle variability, as already observed in the previous characterization of the devices (see Fig.~\ref{fig:cell characteristics}). The presented results do not overlap the two states, so the correct logical output is ensured. Still, the cycle-to-cycle variability could potentially endanger the correctness of the results for this logic type.

In the cases where no switching is allowed, the initial and output states are identical, as expected. These non-switching cases are shown more precisely in Fig.~\ref{fig:results gates} (e). If the cell is initially in the HRS, the resistance remains high. Similarly, the resistance remains low for an initial LRS. Most cases stay constant at a specific resistance, and only case 6 shows variations that HRS instabilities can explain~\cite{Wiefels.2020}. Nevertheless, HRS and LRS regions do not overlap. Case 3 shows minor variations as well since the voltage configuration would also allow a switch, but the initial state $I$ is already '1'. Compared to the HRS instabilities, LRS instabilities are negligible. Only these two cases suffer from this issue since, in all other cases, either the transistor is closed, or there is no voltage difference between TE and BE. In summary, these results ensure correct computation with 1T1R configuration.

\subsection{Scouting Logic}
\begin{figure}[t!]
\centering
\subfloat[]{\scalebox{1}{\includegraphics[width=0.24\textwidth]{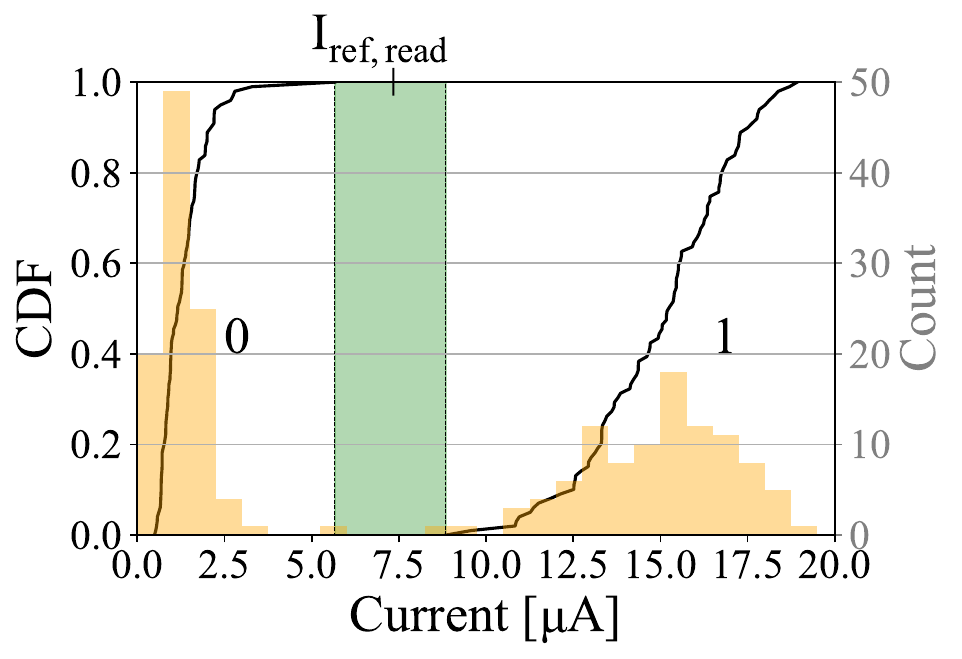}}%
}
\hfill
\subfloat[]{\scalebox{1}{\includegraphics[width=0.24\textwidth]{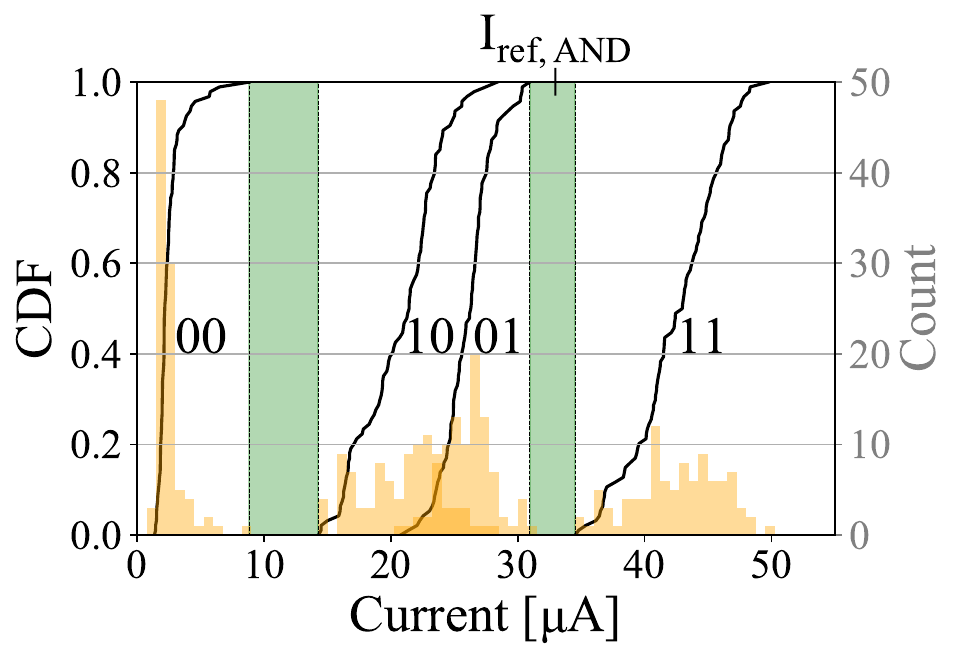}}%
}
\\
\subfloat[]{\scalebox{1}{\includegraphics[width=0.24\textwidth]{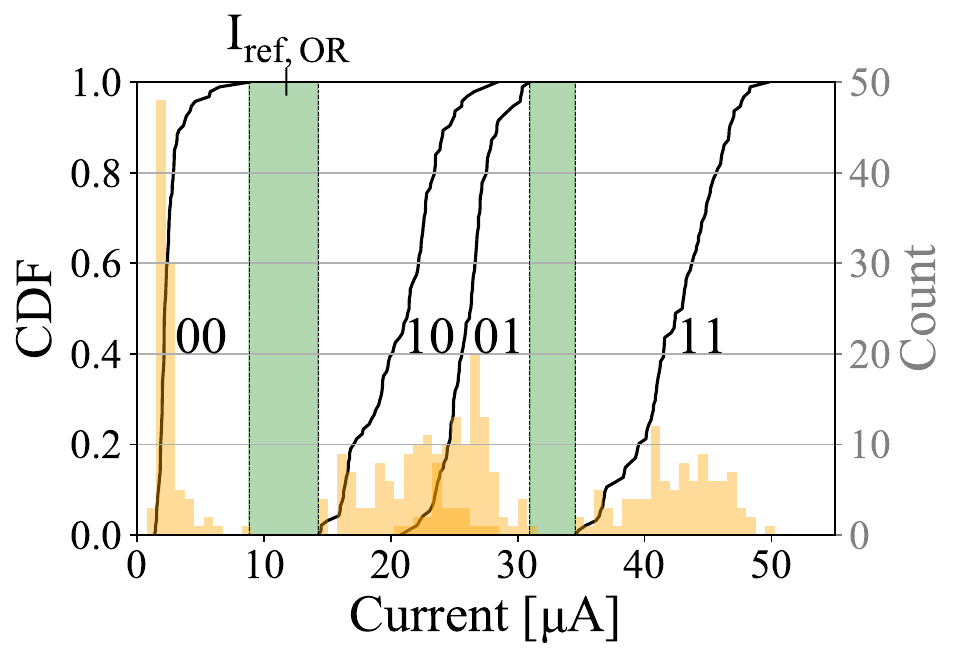}}%
}
\hfill
\subfloat[]{\scalebox{1}{\includegraphics[width=0.24\textwidth]{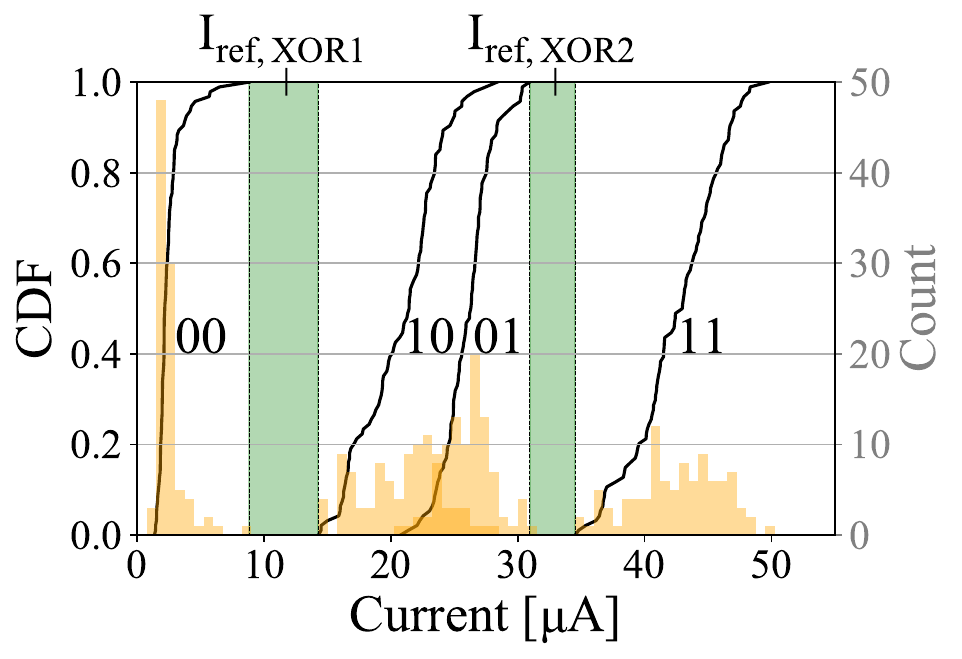}}%
}

\caption{Experimental results for scouting logic with placing of the corresponding reference current between the four input combinations for (a) read, (b) AND, (c) OR, and (d) XOR operations. A CDF plot shows the distribution of currents of 100 cycles for each input combination.}
\label{fig:Scouting_results}
\end{figure}

The results for the scouting logic are shown in Fig.~\ref{fig:Scouting_results}. The gaps between the input combinations make the reference current's reliable placement possible and ensure a correct operation. Therefore, the values \textit{I}\textsubscript{ref,read}= 7.25\,$\upmu$A, \textit{I}\textsubscript{ref,OR}= \textit{I}\textsubscript{ref,XOR1}= 11.55\,$\upmu$A, and \textit{I}\textsubscript{ref,AND}= \textit{I}\textsubscript{ref,XOR2}= 32.74\,$\upmu$A are defined as reference currents for the four functions. For read, AND, and OR, currents higher than the defined reference current correspond to a logical '1', and currents lower than the reference to a logical '0'. For XOR, currents between \textit{I}\textsubscript{ref,XOR1} and \textit{I}\textsubscript{ref,XOR2} represent the output '1', currents outside of these values are considered as '0'.

The cycle-to-cycle variability observed both in the device characterization and the 1T1R logic is visible in the Gaussian shape of the count distribution of the currents (yellow histogram plots in Fig.~\ref{fig:Scouting_results}). If it were even broader, it could result in overlaps of the input states, making a reliable placing of the reference current impossible. Another issue is the cell-to-cell variability, presented in the shift between the black CDF curves of the states 10 and 01 since this shows the different LRS states of different cells. The application of scouting logic can be extended to three or more cells, necessitating a more comprehensive consideration of cell-to-cell variability as the number of cells escalates. This increase can potentially lead to an overlap of input states, thereby increasing the risk of logical failure.

\section{Challenges and Limitations}
\begin{figure}[t!]
\centering
\subfloat[]{\scalebox{0.8}{\includegraphics[width=0.24\textwidth]{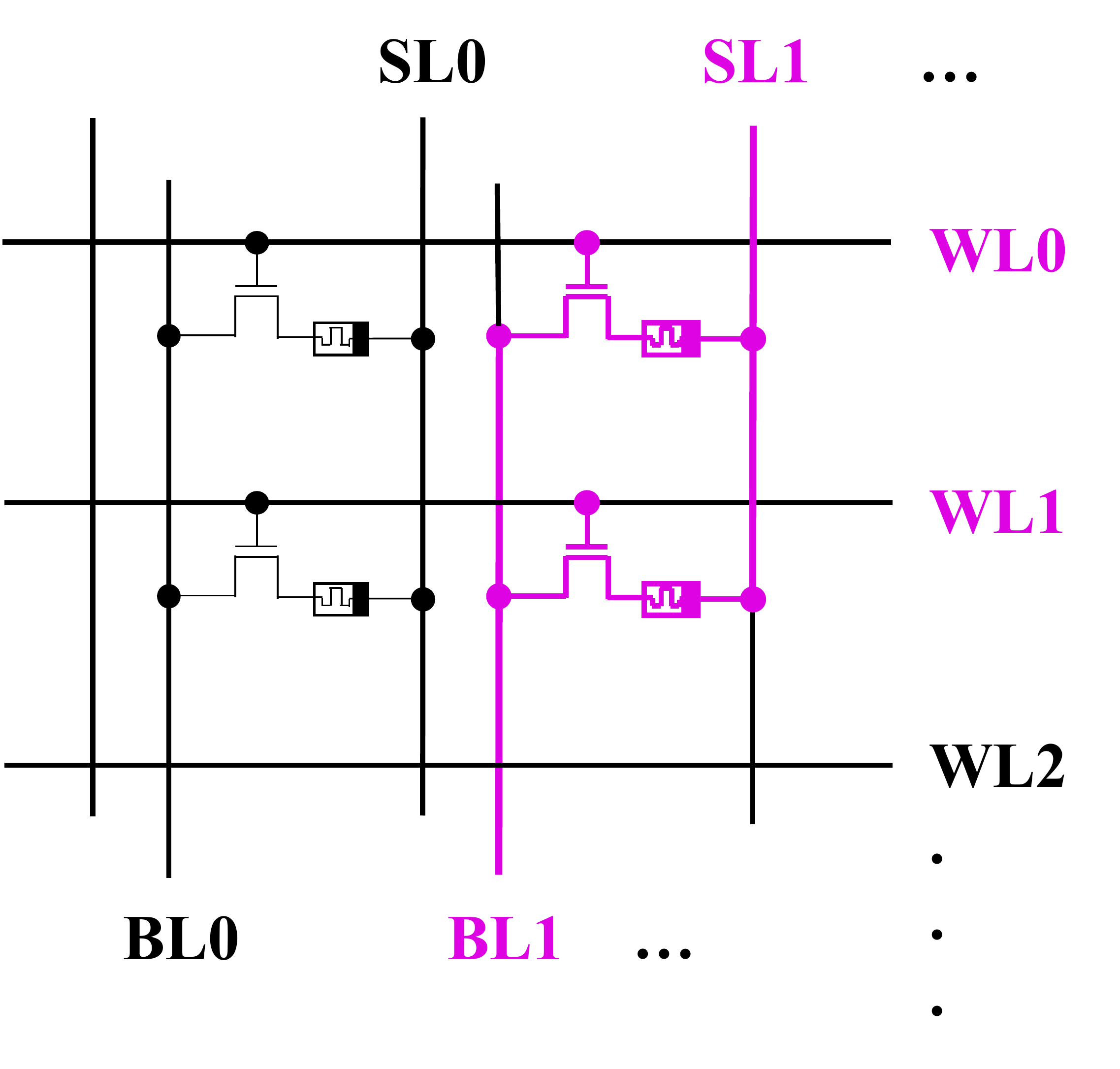}}%
}
\hfil
\subfloat[]{\scalebox{0.8}{\includegraphics[width=0.24\textwidth]{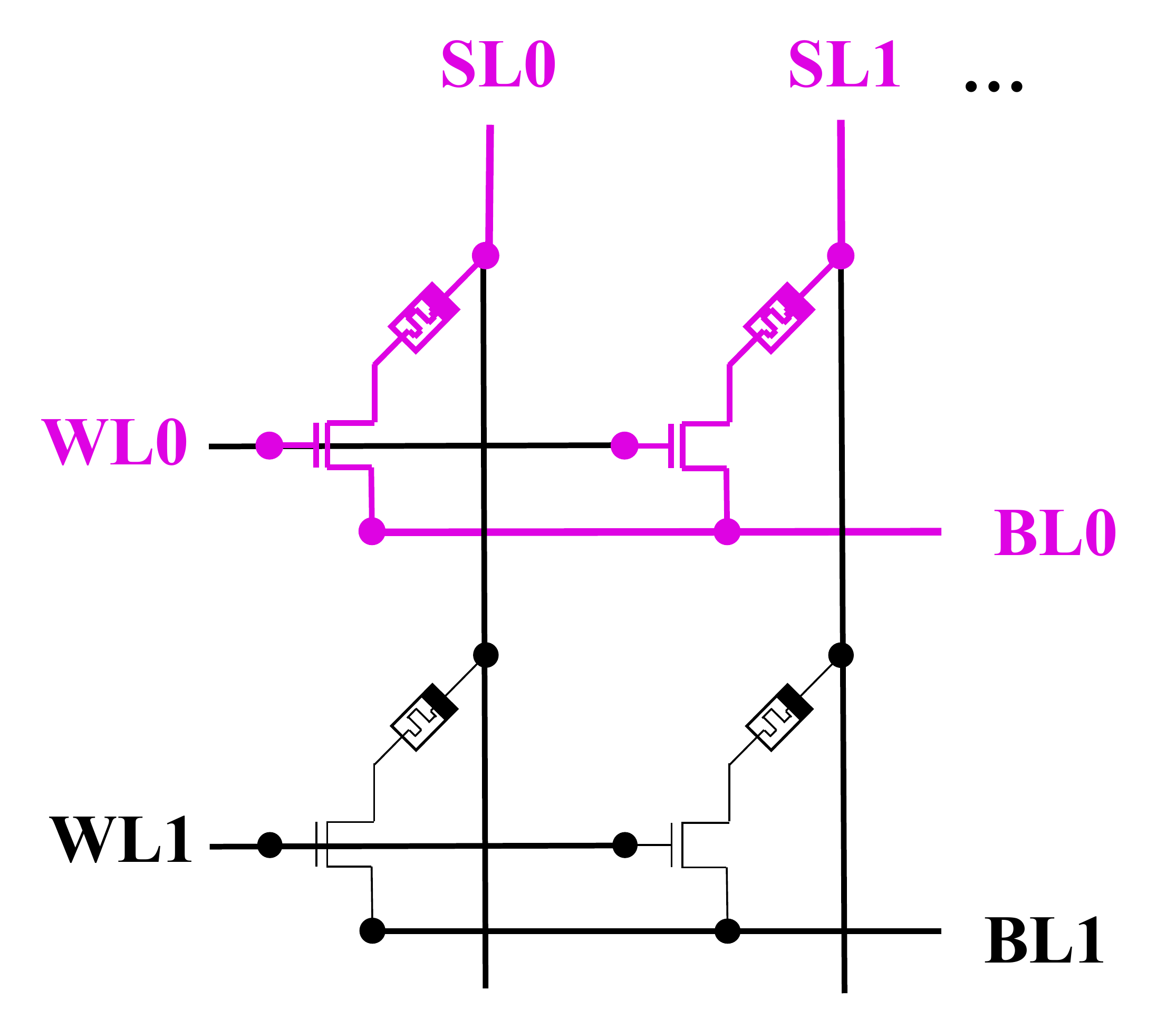}}%
}

        \caption{Connection of two cells in parallel in an (a) 1T1R standard array and a (b) pseudo-crossbar array.}
        \label{fig:array_parallel}
\end{figure}

The 1T1R standard configuration is a widely implemented design due to its high selectivity with minimal leakage currents. However, the column-wise SL and BL design faces challenges for serial connection of multiple cells as well as operating different cells in parallel with different voltages. To connect two cells in parallel, both SL and BL have to have the same index, and two additional WL are selected to open the transistors of two cells, as illustrated in Figure~\ref{fig:array_parallel}(a). The same voltage difference between the top and bottom electrodes is applied to both cells. Since there is no other option for connecting two cells, the application of different voltages at the two cells is impossible. For the demonstrated 1T1R and scouting logic, the given 1T1R configuration fits perfectly; however, this limitation may reduce the potential for other logic in-memory implementations.

A proposed solution for the issue presented is a pseudo-crossbar array, depicted in Fig.~\ref{fig:array_parallel}(b). This design facilitates parallel connection by selecting two SLs, one BL, and one WL enabling the application of different voltages to each cell.

Besides the array architecture, RRAM technology has challenges, especially cycle-to-cycle and cell-to-cell variability. Our results show that the distribution, especially of the HRS, ranges over an order of magnitude. Another challenging issue is the different SET and RESET voltages required for different cells. Finding the exact parameters to ensure a reliable, correct logic operation and minimal power consumption is crucial. 

The 1T1R logic technique exhibits notable advantages due to its simplistic design, employing only a single memristor. Scouting logic demonstrates significant potential as it employs a low voltage and no switching during logical operations, promising reduced power consumption and prolonged device lifespan. However, the need to initialize the inputs occasionally, as well as the need to switch the devices in 1T1R logic, is a major limitation of non-stateful logic. Similar techniques to those used in Resch et al. \cite{Resch.2023} must be used to increase the lifetime of the devices.

\section{Conclusion}

This brief successfully demonstrated computing by SiO\textsubscript{x} VCM cells in a 1T1R array. The cells can distinguish between the high resistance state (HRS; logical '0') and the low resistance state (LRS; logical '1'). Two non-stateful logic types, a complete Boolean function with 1T1R array and scouting logic, have been experimentally investigated. All critical cases in the Boolean set with 1T1R array have been successfully operated without logical failures. Furthermore, the scouting logic had sufficient room for placing a reference current to distinguish between '0' and '1' for the logic functions AND, OR, and XOR. The limitations of the 1T1R array are evident in the parallel connection of cells and applying different voltages at each cell. Additionally, reliability issues of VCM devices still need to be investigated.

With further research and development, the opportunities of this technology can be further explored, ultimately leading to greater efficiency in time and energy.

\bibliographystyle{IEEEtran}
\bibliography{0_Paper_1T1R_Scouting}

\end{document}